# ADJACENCY MATRIX BASED ENERGY EFFICIENT SCHEDULING USING S-MAC PROTOCOL IN WIRELESS SENSOR NETWORKS

Shweta Singh[1] and Ravindara Bhatt[2]

Department of Computer Science and Engineering, Jaypee University of Information Technology, Waknaghat, Distt. Solan, (H.P) India.

shweta.singh7787@gmail.com
ravibhatt749@gmail.com

## ABSTRACT

*Communication is the main motive in any Networks whether it is Wireless Sensor Network, Ad-Hoc networks, Mobile Networks, Wired Networks, Local Area Network, Metropolitan Area Network, Wireless Area Network etc, hence it must be energy efficient. The main parameters for energy efficient communication are maximizing network lifetime, saving energy at the different nodes, sending the packets in minimum time delay, higher throughput etc. This paper focuses mainly on the energy efficient communication with the help of Adjacency Matrix in the Wireless Sensor Networks. The energy efficient scheduling can be done by putting the idle node in to sleep node so energy at the idle node can be saved. The proposed model in this paper first forms the adjacency matrix and broadcasts the information about the total number of existing nodes with depths to the other nodes in the same cluster from controller node. When every node receives the node information about the other nodes for same cluster they communicate based on the shortest depths and schedules the idle node in to sleep mode for a specific time threshold so energy at the idle nodes can be saved.*

## 1. INTRODUCTION

The Wireless Sensor Network has gained popularity in recent decades. It consists of the nodes which are battery operated devices, processors and sensors. As nodes are driven by battery operated devices it is obvious the time will come when the battery will get discharged and node will go down permanently [7]. It is not possible to recharge the battery inside the nodes and make the node in the running state again. Once its battery gets discharged then that node will go down permanently and that node is to be replaced with the other working node. That's why in the design of the Wireless Sensor Network it is very crucial to save the energy at every node by putting the node in to sleep state when the node is idle. As the nodes contains the processor and sensors also with battery the state transition of the processor and sensors takes negligible energy. The nodes are consuming 60 mW, 45 mW, 45 mW, 90 µmW of energy while transmission, reception, listening and sleeping periods respectively [1].When there are several nodes present inside the WSN then it is more suitable to form those nodes inside one cluster so all the nodes inside the cluster are easy to manage rather than to manage each and every node individually. Cluster is the main piece of cake which communicates with the third party. This third party may be base node, controller node, another cluster head, router etc [4][5]. Here as written the energy consumption during the sleep time is 90 µmW so it is better to put the idle node into sleep mode to minimize the power consumption and enhances the battery life [8][13].When the link is shared between the various nodes at this time at a time only one node has the authority to send the data to the other node for that it must be made sure that the channel or link is free or not. This can be done with the help of sending the request to send signal referred as rts to the receiver node and get back clear to send referred as cts from the receiver side to the sender [14]. For this kind exchanging the signals between two nodes both the nodes has to communicate with each other. By doing this kind of communication every node may get information about their neighbour nodes with which the particular node is connected directly. This process of communication between two nodes of exchanging information is referred as Local Gossips [15]. This paper proposes the scheduling algorithm of the nodes by putting them in to sleep mode when they are idle based on information generated by the adjacency matrix and the energy consumption is to be considered during the time of transmitting and receiving only, not during the idle listening at the time of communication based on the shortest depth. This paper is organized as follows. The Section II explains the related work, Section III focuses the proposed model, and Section IV discusses the advantages of the proposed model and section V finally reaches to the Conclusion.

## 2. RELATED WORK

In [2] authors have proposed the clique based node scheduling in which the clique is formed and then the ID assigned with the clique. When the nodes are connected or formed together for performing specific task like communication then the group of nodes are referred as clique. In clique based node scheduling the authors have discussed the sensing range and communication range of the nodes inside the cliques. The nodes with the closure distance are formed into one clique. These different cliques form together forms a network of clique groups. After assigning the group ID to cliques the connectivity and coverage maintenance scheduling is applied with the help of the mathematical equation. During the assignment of the local ID the union of neighbour ID and nearest neighbour ID is performed. Thus the nodes have complete information about the existing nodes inside the clique with in Wireless Sensor Network. In [3] authors have focused mainly on the Multiple Target Coverage Problem (MTCP) in which the authors have discussed that the sensor nodes having the same sensing range consume the same amount of energy. There exist the overlapped targets in sensor networks which are monitored by more than one sensor nodes so the energy is being wasted in monitoring the overlapped targets. That's why only one node must be there to monitor the overlapped target for saving the energy which was wasting at another node. The authors have focused on area coverage algorithm and target coverage algorithm. Area Coverage algorithm maintains both connectivity and full area coverage with a very small communication overhead. In the Target Coverage problem the goal is to monitor all the targets. In [6] the ZigBee network the network consists of the co-ordinator node, some finite number of routers followed by the child router and end devices connected with the routers and child routers directly. The co-ordinator node is responsible for the assignment of the address to the devices. The authors have proposed simple and efficient routing schemes for a Long-thin network with the address assignment scheme and tree routing scheme. In [8] the authors have discussed the sense sleep tree approach. This sense sleep tree approach mainly focuses on the various issues like application specific nature of WSN, Co-ordinated Sleep Scheduling issue, near connected domain partition, spanning tree structure, centralized approach and cross layer design. In [10] the authors have proposed the coverage problem of the nodes. When the nodes inside the wireless sensor network having the largest coverage area more energy is wasted. The authors have focused on the coverage redundancy problem where the locations of nodes and the distances between nodes are not easily calculated. The authors in [10] have defined a neighbour graph as the graph formed by the neighbours of a node. In [11] the authors have enlightened the Linear Distance Based Scheduling (LDS) scheme in which the sensor nodes are placed in to clusters and every cluster has cluster head. The sensor node that is farther away from cluster head has the higher probability to be put in to sleep mode.

The authors have compared LDS performance with Randomized Scheduling scheme.LDS scheme results in longer network lifetime than the RS scheme. In [12] the authors have focuses mainly on the energy efficiency coverage preserving node scheduling in wireless sensor network. This scheduling mainly classified in to two phase namely initial scheduling phase and wake up scheduling phase. Initial scheduling phase is again classified in to two categories which are neighbour information and redundancy evaluation. In neighbour information every node collects the information about the neighbour. In redundancy evaluation scheme as title says the redundant nodes are detected. In the wake up scheduling phase the triggering event occurs when the energy at every node reaches to specific time threshold. According to that time threshold the sleep and wakeup scheduling should be performed. In [9] the sensor nodes are classified in to forwarding nodes and listening nodes. The proposed model [9] focuses mainly on the synchronization of the nodes while sleep and wake up process. The model uses the beacon frame issued from Co-ordinator node when any node is scheduled to be put in to sleep mode. All nodes can be entered in to sleep mode at the same time. The authors have concentrated mainly on the synchronization mechanism required to send the nodes in the sleep mode. Our proposed model mainly focuses on the energy efficient model based on the adjacency matrix. In the proposed model every node, after assignment of ID, collects its neighbour information and calculates the depth. After collecting the neighbour information every node sends the neighbour information ID wise to the controller node via the router. The controller node forms the adjacency matrix and broadcasts the information to the individual clusters via router. When every node gets the information about the existing nodes, the communication with each other node is done based on the shortest depth. The node which is in idle listening is put in to sleep mode for a specific time threshold. In the proposed model the main sender node and final receiver may or may not be synchronized with each other. This model provides the synchronization flexibility in terms of sleep and wake up scheduling for efficient energy scheme in Wireless Sensor Network.

# 3. PROPOSED WORK

In the proposed model the sleep and wake up scheduling algorithm is designed for energy efficient scheduling in Wireless Sensor Networks based on the adjacency matrix. The network consists of the $2^n$ number of the sensor nodes where n<=3 should be taken in the idle case. Taking n<=3 will provide the ease of understanding the concept in a very fruitful manner. The small value of n till 3 will help much better in managing the number of different nodes existing in the Wireless Sensor Network. Anyone can take n value greater than 3. Here maximum 8 nodes can be there inside the Wireless Sensor Networks. These nodes are plugged into a cluster and cluster is responsible for connection with the controller node via the router. Inside the one cluster each node can be connected with any number of nodes. The proposed model uses the S-MAC protocol for local gossiping of the nodes, converge cast message from cluster to different particular nodes and broadcast the message from the controller node to the routers if there are more than one routers connected with the controller node. In the proposed model there can be following scenarios. In first scenario multiple clusters can be connected to the single router. In second scenario single clusters can be connected to an individual dedicated router to that cluster. In both the scenarios the routers will be connected to the single controller node for communication which is shown in figure 1. The proposed scheme is explained in various phases as explained follows.

## 3.1. Assignment of the node ID

In this phase as per discussion there are maximum 8 nodes possible inside one cluster so the nodes should be assigned unique ID in random manner. Here the nodes are given the ID based on the binary numbers starting of the 0000 to 1111. In the proposed structure there exists the case where there are two clusters existing. Each cluster containing six nodes connected with each other. Every Cluster is connected with the router having ID 1000 and the router is connected with the Controller Node having ID 1001. As shown in figure 1 there are two clusters existing referred as Cluster-1 and Cluster-2. In Cluster 1 the nodes are connected with one another named A-B-C-D-E-F and the IDs are assigned to them in the order 0100,0001,0111,0101,0011,0010 respectively, while inside the Cluster 2 the nodes are connected with one another named G-H-I-J-K-L having IDs 1010,1100,1011,1111,1110,1101 respectively. However clusters may have their separate dedicated router connected to the controller node. This has been shown by mentioning cluster N with its own cluster connected to the controller node. During the assignment of IDs to the nodes inside the clusters the IDs of the Router and Controller Nodes which are 1000 and 1001 are fed initially as Current Destination node ID and Final Destination node ID respectively which will be used for future reference as shown in table 1 and table 2.

Table 1. Initial Information fed to the nodes inside Cluster 1

| Cluster Number | Node | Node ID | Destination Node ID | Final Destination Node ID |
|---|---|---|---|---|
| 1 | B | 0001 | 1000 | 1001 |
| 1 | F | 0010 | 1000 | 1001 |
| 1 | E | 0011 | 1000 | 1001 |
| 1 | A | 0100 | 1000 | 1001 |
| 1 | D | 0101 | 1000 | 1001 |
| 1 | C | 0111 | 1000 | 1001 |

Table 2. Initial Information fed to the nodes inside Cluster 2

| Cluster Number | Node | Node ID | Current Destination Node ID | Final Destination Node ID |
|---|---|---|---|---|
| 2 | G | 1010 | 1000 | 1001 |
| 2 | I | 1011 | 1000 | 1001 |
| 2 | H | 1100 | 1000 | 1001 |
| 2 | L | 1101 | 1000 | 1001 |
| 2 | K | 1110 | 1000 | 1001 |
| 2 | J | 1111 | 1000 | 1001 |

## 3.2. Collecting the Nearest Node Information

In this phase every node will have same communication range r. The nodes which are being overlapped in r are referred as the neighbour node of the sensing node. In [5] the authors have suggested about the coverage constraint. The router is situated at the location where it can be overlapped in the communication ranges of the different nodes inside the cluster. With the help of this policy every nodes can transmit their information to the router directly. Every node inside the same clusters tries to collect the nearest neighbour information by checking the nodes which overlap within their communication range and the ID of the nearest neighbour is collected by the nodes. This process is known as local gossips as it is done in the S-MAC protocol of Wireless Sensor Networks. As shown in the figure 1 the nodes inside the Cluster-1 will try to collect the nearest neighbour information based on ID. The node having the smallest ID will be given first preference for collecting the neighbour node information. In the Cluster 1 node B will be given first chance for gathering its neighbour information which are node A and node C having IDs 0100 and 0111 respectively. Similarly after node B the other nodes inside Cluster 1 will be F, E, A, D, C in the ascending orders of lowest IDs respectively. However this priority can be changed in the next round randomly so every node can get higher priority once.

### 3.3. Calculating the Depths

In this phase when every node inside the same cluster finishes collecting the information regarding its neighbour nodes, they will start calculating the distance/depth between two nodes. The Depth between the two nodes is referred as the longest available path from one node to another node. This Depth Calculation follows the given formulae. Depth = (largest ID of a node – smallest ID of a node). In the case of the Cluster 1 the depth between node A and node B is calculated as (node A's ID- node B's ID) which are 0100-0001 so the answer is 3 similarly the depth between other nodes in the cluster 1 are 6,2,2,1,2 respectively which is explained clearly in the figure 1. Same will be true for the Cluster 2.

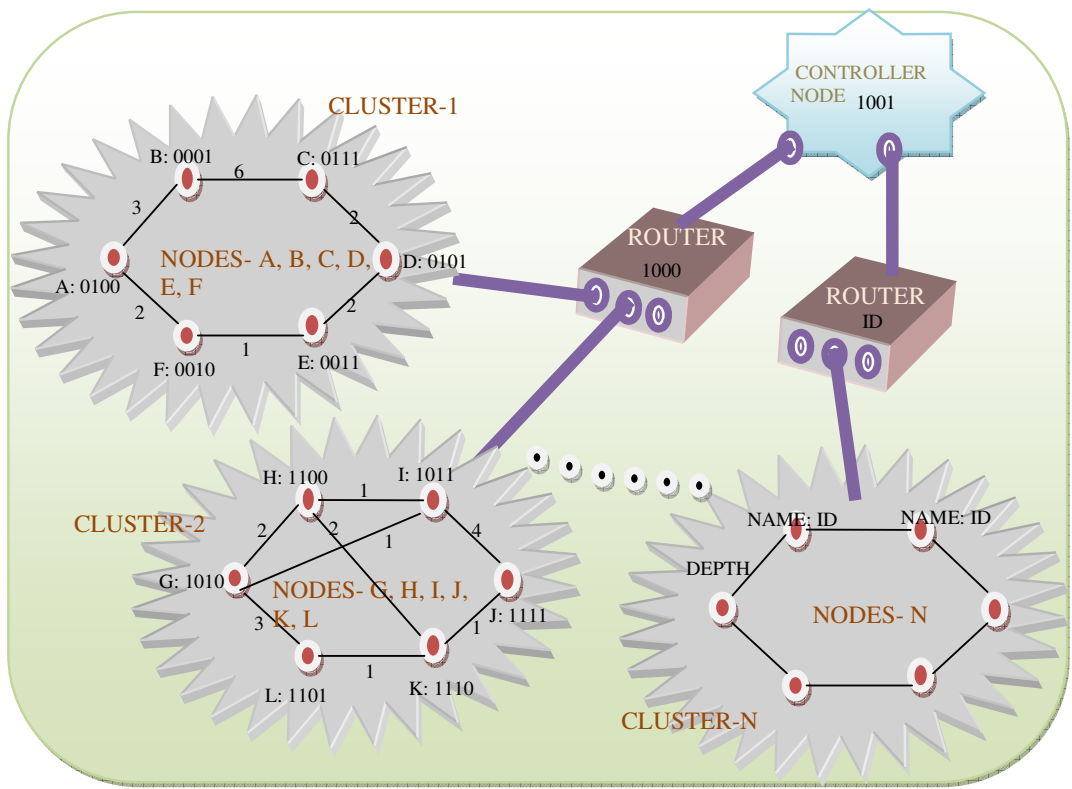

Figure 1. Proposed Architecture having n Clusters, Nodes, Router and Controller Node

### 3.4. Sending the neighbour node's ID information to the Router

In this phase every node inside the same cluster sends the neighbour ID information to the Router having the ID 1000 after local gossips. This process follows the converge cast mechanism because every node sends the neighbour information to the router one by one when their turn comes. As discussed in the phase 1every nodes are fed the IDs of Router and Controller Node during assignment of the IDs to the nodes. Every node inside the same cluster will send its neighbour ID information in the same order as they collect their neighbour information. In the Cluster 1 the node B will be given first preference in sending its neighbour ID information to the Router. After B the other nodes are F, E, A, D, C will send their neighbour ID information to the Router. The Router will receive the information about the Cluster Number, Node, Node ID, IDs of Neighbour Node, Depth of one node with respect to its neighbour node, Current Destination, Final Destination as shown in the table 3 and 4. The Current Destination is the ID of the Router which is 1000 and Final Destination is considered as the ID of the Controller Node which is 1001. According to the table 3 and table 4 the Router will send the neighbour ID information Cluster wise and ID wise to the Controller Node one by one. Thus Controller node will have the information about the different Cluster's node Information from the Router. Converge cast mechanism provides the fruitful way in collecting the node information cluster wise to the controller node via router.

Table 3.  Node Information of Cluster 1 at Router with ID 1000

| Cluster Number | Node | Node ID | IDs of Neighbour Node | Depth | Current Destination | Final Destination |
|---|---|---|---|---|---|---|
| 1 | B | 0001 | 0100,0111 | 3,6 | 1000 | 1001 |
| 1 | F | 0010 | 0100,0011 | 2,1 | 1000 | 1001 |
| 1 | E | 0011 | 0010,0101 | 1,2 | 1000 | 1001 |
| 1 | A | 0100 | 0001,0010 | 3,2 | 1000 | 1001 |
| 1 | D | 0101 | 0011,0111 | 2,2 | 1000 | 1001 |
| 1 | C | 0111 | 0001,0101 | 6,2 | 1000 | 1001 |

Table 4.  Node Information of Cluster 2 at Router with ID 1000

| Cluster Number | Node | Node ID | IDs of Neighbour Node | Depth | Current Destination | Final Destination |
|---|---|---|---|---|---|---|
| 2 | G | 1010 | 1100,1101,1011 | 2,3,1 | 1000 | 1001 |
| 2 | I | 1011 | 1100,1111,1010 | 1,4,1 | 1000 | 1001 |
| 2 | H | 1100 | 1011,1010,1110 | 1,2,2 | 1000 | 1001 |
| 2 | L | 1101 | 1110,1010 | 1,3 | 1000 | 1001 |
| 2 | K | 1110 | 1111,1101,1100 | 1,3,2 | 1000 | 1001 |
| 2 | J | 1111 | 1011,1110 | 4,1 | 1000 | 1001 |

## 3.5. Gathering the information and Form the Adjacency Matrix at the Controller Node

In this phase the Controller node gathers the neighbour ID information from various nodes residing in the same cluster as well as from different Clusters. As shown in table 5 the Controller node gathers the information from the Router which contains the Cluster Number, Node, Node ID, IDs of Neighbour Node, Source ID, and Destination ID. Table 5 has explained the neighbour node information for cluster 1 and cluster 2. After receiving the information from the router the Controller Node starts forming the Adjacency Matrix ID wise and cluster wise as shown in table 6 and table 7. At the time of forming the adjacency matrix the controller node puts 1 in place of neighbour ID and puts 0 when it does not find the ID of any node inside the neighbour ID list of a particular node. Every time for generating the adjacency matrix the controller node has to refer the table 5 from which it can verify which nodes are neighbours of the particular node. In table 6 the adjacency matrix is explained for cluster 1. After generation of adjacency matrix the controller node will gather the information about the ID of Node, Neighbour Node ID, Not neighbour ID Node, the union of the Neighbour ID and Not Neighbour ID Node, Destination Node ID, Cluster and Depth of Each node with reference to its neighbour node. This is explained in a rich manner in the table 8. Please note that the controller node is doing Union Process of ID with the not neighbour ID of node for getting the information how many nodes are there in a cluster of Wireless Sensor Network. After Union Process the controller node identifies the total number of nodes residing inside the cluster 1. The Controller node will also utilize the depth column as sent by router to it already at the time of broadcasting the node information to the router back.

Table 5.  Cluster Information at Controller node from Router

| Cluster Number | Node | Node ID | IDs of Neighbour Node | Source ID | Destination ID |
|---|---|---|---|---|---|
| 1 | B | 0001 | 0100,0111 | 1000 | 1001 |
| 1 | F | 0010 | 0100,0011 | 1000 | 1001 |
| 1 | E | 0011 | 0010,0101 | 1000 | 1001 |
| 1 | A | 0100 | 0001,0010 | 1000 | 1001 |
| 1 | D | 0101 | 0011,0111 | 1000 | 1001 |
| 1 | C | 0111 | 0001,0101 | 1000 | 1001 |
| 2 | G | 1010 | 1100,1101,1011 | 1000 | 1001 |
| 2 | I | 1011 | 1100,1111,1010 | 1000 | 1001 |
| 2 | H | 1100 | 1011,1010,1110 | 1000 | 1001 |
| 2 | L | 1101 | 1110,1010 | 1000 | 1001 |
| 2 | K | 1110 | 1111,1101,1100 | 1000 | 1001 |
| 2 | J | 1111 | 1011,1110 | 1000 | 1001 |

Table 6. Adjacency Matrix at Controller Node for Cluster 1

| Cluster Number | Node | Node ID | 0001 | 0010 | 0011 | 0100 | 0101 | 0111 |
|---|---|---|---|---|---|---|---|---|
| 1 | B | 0001 | 0 | 0 | 0 | 1 | 0 | 1 |
| 1 | F | 0010 | 0 | 0 | 1 | 1 | 0 | 0 |
| 1 | E | 0011 | 0 | 1 | 0 | 0 | 1 | 0 |
| 1 | A | 0100 | 1 | 1 | 0 | 0 | 0 | 0 |
| 1 | D | 0101 | 0 | 0 | 1 | 0 | 0 | 1 |
| 1 | C | 0111 | 1 | 0 | 0 | 0 | 1 | 0 |

Table 7. Adjacency Matrix at Controller Node for Cluster 2

| Cluster Number | Node | Node ID | 1010 | 1011 | 1100 | 1101 | 1110 | 1111 |
|---|---|---|---|---|---|---|---|---|
| 2 | G | 1010 | 0 | 1 | 1 | 1 | 0 | 0 |
| 2 | I | 1011 | 1 | 0 | 1 | 0 | 0 | 1 |
| 2 | H | 1100 | 1 | 1 | 0 | 0 | 1 | 0 |
| 2 | L | 1101 | 1 | 0 | 0 | 0 | 1 | 0 |
| 2 | K | 1110 | 0 | 0 | 1 | 1 | 0 | 1 |
| 2 | J | 1111 | 0 | 1 | 0 | 0 | 1 | 0 |

### 3.6. Broadcasting the complete node information back to the Router

In this phase the controller node will broadcast the complete node information to the router along with the depth between different nodes. Router will check the destination ID which will be 1000 and Cluster number and match with its own ID. After verifying the ID the router will send the information to the specified Cluster because the router has the information about the nodes inside specified cluster as specified in the table 3 and table 4 for cluster 1 and cluster 2 respectively hence based on those tables the router will send the information to the particular cluster's node ID wise using converge cast mechanism because node to node communication comes in to picture. Here the router will be using Node ID as a destination ID. Thus, after receiving the information from the router the nodes of the particular cluster will have the complete information about the existing other nodes inside the same cluster of Wireless Sensor Network.

### 3.7. Sleep and Wake up scheduling procedure while communication in the same cluster

In this phase every nodes inside cluster 1 has information about its neighbour node and the other existing nodes inside the same cluster. The scheduling mechanism is to be implemented in real manner in this phase. Every node will be given specific time threshold based on the priority in the same order as they were passing their neighbour information to the router. In this time threshold only one node has authority to send the data to other nodes and send rts signal as a main sender to other nodes while other remaining nodes will be working as supporting nodes for passing the message of one sender to the final destination receiver node till the specific time threshold for communication is getting completed. This priority will change dynamically ID wise so according to fair share concept every node can get highest priority once but initially the lowest node ID will be having highest priority for communication. After the priority moves from node B to node F then node E then node A when node A's turn comes what happens is explained in very fruitful manner. The scheduling mechanism will work as follows. Let us say node A having ID (0100) wants to communicate or send some data frames to the node D having ID (0101) then firstly it must be decided at which way the data frames to be sent and how the scheduling should be performed. Here according to table 8 the ID of Node, its neighbour ID & depth of that node towards its neighbours are referred. In the depth column in table 8 there are two neighbours of every node. The depths are indicated by depth of node A to node B, depth of node C to node B likewise indicated. In depth column the shortest depth to one node from anther node that node will be selected for transmitting the data frames and other nodes with highest depth will be put in to sleep mode, so in the proposed case as for example A (0100) wants to send some data frames to D (0101) then according to depth column presented in table 8 node F with ID 0010 and depth 2 will be selected and node A will first send rts (request to send) signal to node F. When node F receives rts signal it will send cts (Clear to send) signal to node A so it is understood that there is no link failure between node A to node F. Having received the cts signal from node F node B which is the highest depth neighbour of node A will be scheduled to be put in to sleep mode. Now node A sends some data frame to node F it will pass the destination node's ID which is node D (0101) ,current source ID which is node A (0100) and current destination ID which is node F (0010) and waits for the acknowledgement. When node F receives the data it will read destination node's ID and gives acknowledgement back to node A. In reply node A sends confirmation to the node F that it has received the acknowledgement successfully and scheduling is done to be put node A in to sleep mode for certain time threshold say 6 milliseconds. Thus three way handshaking mechanism is used, after this the node F again refer depth column in table 8 and selects the shortest depth excluding sender's depth which is 1 and sends the rts signal to node E and wait till node E sends it back cts signal. When node E sends the cts signal to node F then it sends data frames toward node E (0011). The node F will follow the same mechanism as node A was following towards node E and waits for acknowledgement. When node F gets the acknowledgement from node E it will send reply in answer to acknowledgement and will be put in to sleep mode for a time threshold say 4 milliseconds. Same will be scenario with the node E which will send data frames to node D and waits for the acknowledgement. When node E gets acknowledgement from node D it will give reply to node D that it has received the acknowledgement successfully. Now node E is not sent to sleep node because when node D receives the data frame and proceed with the data frames the nodes which had been put in to sleep mode will complete their time threshold so the nodes who were sleeping are node A, B, C, F thus they all are awaken because their time threshold will get completed till the last node receives and acknowledges the data frames. Thus in the proposed scheduling it is possible to save the energy at nodes A, B, C, F in idle listening. All the nodes are now awaken up for the next transmission thus the process will get repeated throughout the cluster inside the wireless sensor networks. In the specified scenario as shown in figure 1 the data frames are sent to node D from node A via the path A-F-E-D instead of A-B-C-D. The priority based time threshold assigned to every node makes sure that with in one specific time threshold only one communication will be performed so no other nodes with in node A's time threshold will send rts signal to any other node as a main sender because every node has to wait till their turn comes for becoming main sender. Please note that propose model focuses mainly on energy saving by putting idle node into sleep mode not focuses on the routing of the data frames. The data frames may travel via longest depth in case of link failure exists on the shortest depth but the energy saving at the node is most important and the main cup of tea in the proposed paper. This whole process is explained via the scheduling algorithm which is explained as follows.

Table 8. Complete Node Information with Depths at Controller Node for Cluster 1.

| Node | Node ID | Neighbour Node ID | Not Neighbour ID | Neighbour ID U Not Neighbour ID | Destin -ation | Cluster Number | Depth |
|------|---------|-------------------|------------------|--------------------------------|---------------|----------------|-------|
| B | 0001 | 0100,0111 | 0001,0010,0011, 0101 | 0001,0010,0011,0100,0101,0111 | 1000 | 1 | 3,6 |
| F | 0010 | 0100,0011 | 0001,0010,0101, 0111 | 0001,0010,0011,0100,0101,0111 | 1000 | 1 | 2,1 |
| E | 0011 | 0010,0101 | 0001,0011,0100, 0111 | 0001,0010,0011,0100,0101,0111 | 1000 | 1 | 1,2 |
| A | 0100 | 0001,0010 | 0011,0100,0101, 0111 | 0001,0010,0011,0100,0101,0111 | 1000 | 1 | 3,2 |
| D | 0101 | 0011,0111 | 0001,0010,0100, 0101 | 0001,0010,0011,0100,0101,0111 | 1000 | 1 | 2,2 |
| C | 0111 | 0001,0101 | 0010,0011,0100, 0111 | 0001,0010,0011,0100,0101,0111 | 1000 | 1 | 6,2 |

## 3.8. Sleep and Wake up node scheduling Algorithm

Initially it is assumed that all nodes listen the channel in idle mode and with the help of the router every nodes will have the information about other existing nodes. All nodes have been assigned priority same as they were sending their neighbour information to the router and specific constant time threshold has been allocated to every node initially.

Table 9. Scheduling Algorithm

| Steps | Procedure |
|-------|-----------|
| Step1 | Determine Main Sender's ID and Final Receiver's ID where data has to be sent based on priority and specific time threshold has been assigned to them. |
| Step 2 | Find Main Sender's Neighbour node ID with the depth as shown in table V. |
| Step 3 | Select the shortest depth of the node among all the depths for transmitting rts signal. |
| Step 4 | Send the rts (Request to Send)signal to shortest depth node |
| Step 5 | Waits for the cts signal as acknowledgement signal |
| Step 6 | IF acknowledgement = = cts (Clear to Send) signal then |
| Step 7 | Put highest depth node in to sleep mode for specific time threshold and go to step 20 |
| Step 8 | Else |
| Step 9 | Waits for cts as acknowledgement till specific time threshold |
| Step 10 | IF acknowledgement = = cts after some time threshold then go to step 20 |

| Step 11 | Else |
| --- | --- |
| Step 12 | There is a link failure in the network |
| Step 13 | Send rts signal to the longest depth node and waits for cts signal as acknowledgement |
| Step 14 | If acknowledgement = = cts signal then go to step 20 |
| Step 15 | Else |
| Step 16 | There is link failure at every side of neighbours and terminate the communication of the main sender node and go to step 45 |
| Step 17 | End IF |
| Step 18 | End IF |
| Step 19 | End IF |
| Step 20 | While Current Destination ID! = Final Destination ID |
| Step 21 | Send the Data frames to the shortest depth nodes by sending Main Source ID, Current Source ID, Destination ID and Final Destination ID. |
| Step 22 | Wait for Acknowledgement for specified time threshold |
| Step 23 | IF Current Destination node receives data frames successfully then |
| Step 24 | Send acknowledgement signal to the current source ID |
| Step 25 | Else |
| Step 26 | Wait for data frames till complete reception. |
| Step 27 | End IF |
| Step 28 | IF Current Source Node gets acknowledgement signal successfully then |
| Step 29 | Send Acknowledgement to current destination for reception of acknowledgement signal successfully. |
| Step 30 | Put the current source node to sleep mode. |
| Step 31 | Else |
| Step 32 | Resend the data frames to the current destination and waits for the acknowledgement. |
| Step 33 | End IF |
| Step 34 | Read Source ID, Current Source ID, Current Destination ID, Final Destination ID at current destination node. |
| Step 35 | Go to Step 3 |
| Step 36 | End While |
| Step 37 | IF Current Destination ID = = Final Destination ID then |
| Step 38 | Send acknowledgement to the current source ID by Final Destination ID. |
| Step 39 | Send the confirmation of reception of acknowledgement signal to Final Destination ID by the current Source Node. |
| Step 40 | While there are still transmission/reception exists in the cluster |
| Step 41 | Awake the sleeping nodes for next transmission/reception after completion of specific time threshold. |
| Step 42 | Change the priority assigned to the nodes initially in random manner and don't change the specific constant time threshold allocated to the ID initially |
| Step 43 | Go to step 1 |
| Step 44 | End While |
| Step 45 | End of the Algorithm. |

In the sensor network where nodes are dynamic the scheduling algorithm will work as explained in table 9. However there will be little change in the allotment of time threshold T of cluster head. T will start after step 41 in the algorithm. During T every process regarding sending 0 bit, adding newly added node to the cluster is performed by nodes and by the broadcasting process to the nodes by the cluster head is performed. When every process finishes step 42 of the algorithm will get executed. The scheduling will be done by just adding small change in the algorithm.

### 3.9. Changes in the adjacency matrix for dynamic nodes inside the cluster

In wireless Sensor network the node movement is the most sensitive issue which must be resolved very precisely. If the nodes inside the clusters are movable, the structure of the system will include one cluster head. The system should be designed in such a way that all the nodes with the communication range r will overlap the cluster head. With the help of this every nodes are directly connected to the cluster head. In initial phase every node will pass their neighbour information to the cluster head the ID of which is fed to the nodes initially instead of router ID and the ID of the router is fed to the cluster head. The cluster head will collect the information from every node of the cluster about the depths and their neighbour ID and passes the information to the router. The router will pass the information to the controller node. The controller node will proceed same as in the static network and pass the information to the router by adding the column named as bit 0 node ID list and from the router the information will be supplied to the cluster head which is shown in table 12. The proposed system architecture is shown in figure 2. In this dynamic network where every node intend to move from one location to other location the cluster head will be responsible for the maintaining the information in to its buffer and broadcast the information about the other existing nodes to the connected nodes inside the particular cluster. The scheduling of the sensor nodes will be done same as in the static network.

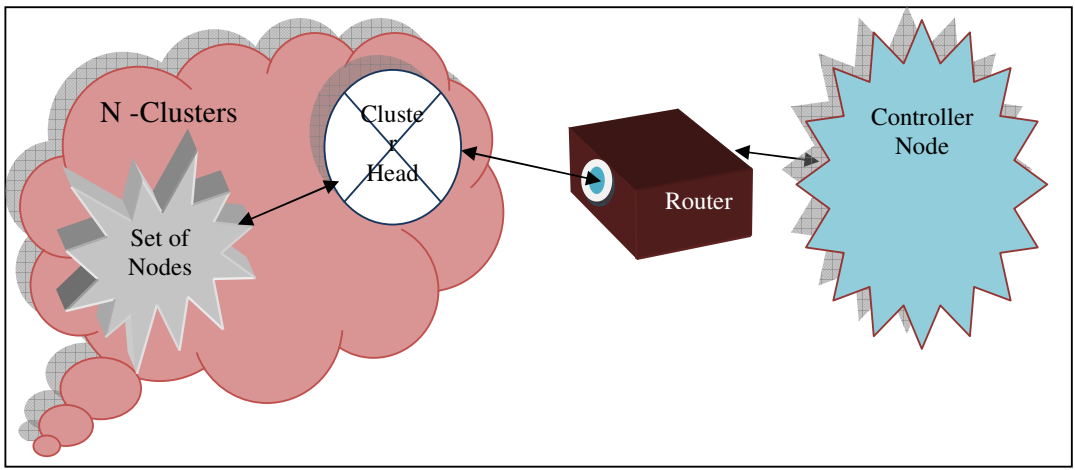

Figure 2. Proposed Architecture having n Clusters, Set of Nodes, Cluster Head, Router and Controller Node in the Dynamic Sensor network where node movement is allowed.

The dynamic sensor network differs compared to static sensor network in following cases.

Case (a). The node wants to move outside the cluster.

The cluster head will always remain in an active state, so any node which wants to move outside to the cluster can send the bit 0 to the cluster head. The node will be allowed to leave the cluster on every (n+1)th transmission by the cluster head. Thus node has to inform to the cluster head during nth time threshold that it going to leave the cluster. Keeping the cluster head always active facilitates any node to move during any time threshold. According to the static node scheduling, inside the cluster every node will have fix time threshold in which the node behaves as a main sender to communicate with any node in the cluster. When first time threshold completes, the system is designed in such a way that if a node wants to leave the cluster in next time threshold then it has to inform to the cluster head in the next time threshold it is going to leave the cluster by sending the bit 0. On completion of first time threshold of a node the time threshold is allocated to the cluster head during which it can receive the bit 0 from the dynamic nodes which wants to move. The cluster head will wait for the bit 0 from the dynamic nodes till its time threshold completes. If it receives the bit 0 from any node with in allocated time threshold it updates its information table by putting the node ID into bit 0 list which it had received from the router and broadcasts the received ID of the node from bit 0 node list to every nodes existing inside the cluster excluding the movable node. Cluster head itself removes the ID of movable node from the neighbour node ID list, not neighbour node ID list of the nodes and Neighbour ID union Not Neighbour ID and puts that ID to bit 0 node list which is shown in table 12. As shown in table 12 the depth of the node is also being cancelled by the cluster head. This changed information table is broadcasted to the nodes. Thus, every node can update their table and try to avoid the communication with the bit 0 node. However if movable node wants to communicate with any node of the cluster within its specified time threshold, it can communicate with the other node and nodes will be scheduled same as in the static networks. If cluster head doesn't receive the bit 0 from any node, it does nothing and the nodes communicate with each other same as in the static network doing and will be scheduled accordingly for saving the energy. Table 11 explains the node A is sending the bit 0 signal to the cluster head.

Table 10 Cluster head with ID 10000 have the information table supplied by the controller node via router

| Node | Node ID | Neighbour Node ID | Not Neighbour ID | Neighbour ID U Not Neighbour ID | Destin-ation | Cluster Number | Depth | Bit 0 node list |
|------|---------|-------------------|------------------|--------------------------------|--------------|----------------|-------|-----------------|
| B | 0001 | 0100,0111 | 0001,0010,0011,0101 | 0001,0010,0011,0100,0101,0111 | 1000 | 1 | 3,6 | - |
| F | 0010 | 0100,0011 | 0001,0010,0101,0111 | 0001,0010,0011,0100,0101,0111 | 1000 | 1 | 2,1 | - |
| E | 0011 | 0010,0101 | 0001,0011,0100,0111 | 0001,0010,0011,0100,0101,0111 | 1000 | 1 | 1,2 | - |
| A | 0100 | 0001,0010 | 0011,0100,0101,0111 | 0001,0010,0011,0100,0101,0111 | 1000 | 1 | 3,2 | - |
| D | 0101 | 0011,0111 | 0001,0010,0100,0101 | 0001,0010,0011,0100,0101,0111 | 1000 | 1 | 2,2 | - |
| C | 0111 | 0001,0101 | 0010,0011,0100,0111 | 0001,0010,0011,0100,0101,0111 | 1000 | 1 | 6,2 | - |

Table 11 Node A (0100) sends the bit 0 information to the cluster head (ID 10000)

| Node | Node ID | Bit | Destination ID |
|------|---------|-----|----------------|
| A | 0100 | 0 | 10000 |

Table 12 Updated information table by the cluster head broadcasted to the nodes inside the cluster

| Node | Node ID | Neighbour Node ID | Not Neighbour ID | Neighbour ID U Not Neighbour ID | Destin-ation | Cluster Number | Depth | Bit 0 node List |
|------|---------|-------------------|------------------|----------------------------------|--------------|----------------|-------|-----------------|
| B | 0001 | 0111 | 0001,0010,0011,0101 | 0001,0010,0011,0101,0111 | 10000 | 1 | 6 | 0100 |
| F | 0010 | 0011 | 0001,0010,0101,0111 | 0001,0010,0011,0101,0111 | 10000 | 1 | 1 | 0100 |
| E | 0011 | 0010,0101 | 0001,0011,0111 | 0001,0010,0011,0101,0111 | 10000 | 1 | 1,2 | 0100 |
| D | 0101 | 0011,0111 | 0001,0010,0101 | 0001,0010,0011,0101,0111 | 10000 | 1 | 2,2 | 0100 |
| C | 0111 | 0001,0101 | 0010,0011,0111 | 0001,0010,0011,0101,0111 | 10000 | 1 | 6,2 | 0100 |

Case (b) Node added to the cluster coming from the other cluster.

The node which is coming to one cluster from the other cluster, it sets itself to some location and on successful setting up its position it tries to broadcast its ID to its full communication range. If the communication range of newly coming node overlaps to the existing nodes inside the cluster and if that node is active, both the nodes will be doing local gossips and exchange their ID information and depth. During the time threshold of the cluster head the nodes whose communication range overlaps with the range of newly added node to the cluster, send their updated neighbour list with ID to the cluster head with depths. Cluster head will update its information table by adding the new neighbour node ID to the sender nodes and add the depth to the depth column, creates new entry of newly added node with forming every basic information in its table like node, node ID, neighbour node ID, not neighbour ID, neighbour ID unions not neighbour ID, Destination address (Cluster Head's Address), cluster number, depth, bit 0 node list. This updated table is broadcasted to every node including the sender nodes. In this way, the existing nodes succeed in getting the information about newly added nodes to the cluster. The newly added node gets the information about the whole cluster's existing nodes by its neighbour node with in one of its active time. The newly added node will get success in collecting the existing nodes inside the network. The same constant time threshold is allocated to the newly added node and priority to become this new node as main sender may get changing in random manner. On finishing the time threshold of the cluster head the next node will become main sender according to the priority and next time threshold and communication will occur among the nodes including the newly added node. The scheduling mechanism in this case will be same as in the static network. The overlapped node sends the information to the cluster head as explained in table 13 and sends the previous information table after updating the neighbour ID to the newly added node as explained in table 14. Suppose node G with ID from cluster 2 is moving to cluster 1 and overlaps with the communication range of node B. Node B will send the information to the cluster head following information. This update is maintained by cluster head and broadcasted to every node excluding the sender node.

Table 13. Node B (0001) sends the cluster head to the updated information about newly added node G (1010)

| Node | Node ID | Neighbour Node ID | Not Neighbour ID | Neighbour ID U Not Neighbour ID | Destination | Cluster Number | Depth | Bit 0 node List |
|------|---------|-------------------|------------------|-------------------------------|-------------|----------------|-------|------------------|
| B | 0001 | 0100,0111,**1010** | 0001,0010,0011, 0101 | 0001,0010,0011,0100,0101,0111,**1010** | 10000 | 1 | 3,6,**9** | - |

Table 14. Cluster head sends the above updated table to every nodes and node B will send the information to the newly added node G.

| Node | Node ID | Neighbour Node ID | Not Neighbour ID | Neighbour ID U Not Neighbour ID | Destination | Cluster Number | Depth | Bit 0 node List |
|------|---------|-------------------|------------------|-------------------------------|-------------|----------------|-------|------------------|
| B | 0001 | 0100,0111,1010 | 0001,0010,0011,0101,1010 | 0001,0010,0011,0100,0101,0111,1010 | 10000 | 1 | 3,6,9 | - |
| F | 0010 | 0100,0001 | 0001,0010,0101,0011,1010 | 0001,0010,0011,0100,0101,0111,1010 | 10000 | 1 | 2,1 | - |
| E | 0011 | 0010,0101 | 0001,0011,0100,0011,1010 | 0001,0010,0011,0100,0101,0111,1010 | 10000 | 1 | 1,2 | - |
| A | 0100 | 0001,0010 | 0011,0100,0101,0011,1010 | 0001,0010,0011,0100,0101,0111,1010 | 10000 | 1 | 3,2 | - |
| D | 0101 | 0011,0111 | 0001,0010,0100,0101,1010 | 0001,0010,0011,0100,0101,0111,1010 | 10000 | 1 | 2,2 | - |
| C | 0111 | 0001,0101 | 0010,0011,0100,0011,1010 | 0001,0010,0011,0100,0101,0111,1010 | 10000 | 1 | 6,2 | - |

| G | 10 10 | 0001 | 0010,001 1,0100,01 01,0111,1 010 | 0001,0010,0011,0 100,0101,0111,10 10 | 10000 | 1 | 9 | - |
|---|---|---|---|---|---|---|---|---|

## 3.10. Energy Consumption at various nodes during Sleep and Wake up Scheduling

In Wireless Sensor networks when two nodes communicate with each other there are four possible states of nodes namely Transmitting, Receiving, Listening and Sleeping. The Energy Cost by the node $v_i$ in all states is given by following equation.

$$[Xi, S, t \times Ptx + Xi, R, t \times Prcv + Xi, L, t \times Plst + Xi, P, t \times Pslp] \times ts$$

Where $X_{i,S,t}$ = indicator whether node $v_i$ Transmits at time t, $X_{i,R,t}$ = indicator whether node $v_i$ Receives at time t, $X_{i,P,t}$ = indicator whether node $v_i$ Sleeps at time t, $X_{i,L,t}$ = indicator whether node $v_i$ Listens at time t, $P_{tx}$ = Energy Consumption in Transmitting, $P_{rcv}$ = Energy Consumption in Receiving, $P_{lst}$ = Energy Consumption in Listening. $P_{slp}$ = Energy Consumption in Sleeping, $t_p$ = time needed to poll channel once, $r_{vi}$ = Data packets per period by vi, $L_{data}$ = Data packets Length, $t_b$ = time to transmit or receive the byte, T= a scheduling period, $t_s$ = slot size, $P_s$ = No of packets transmitted in a slot, k= No of nodes involved in receiving process and idle listening process respectively. According to proposed model the energy cost by node $v_i$ in all states are given by the formula as follows in whole network during scheduling.

$$[Xi, S, t \times Ptx + \{k \times (Xi, R, t \times Prcv + Xi, L, t \times Ptx + (Prts + Pcts))\}] \times ts$$

Where $P_{rts}$= Energy Consumption in sending Request to Send signal, $P_{cts}$= Energy Consumption in sending Clear to Send signal. In the above equation the sleeping states of the transition while communication has been omitted because in sleeping state the node consumes the energy in microamperes. Now how the proposed model fits in to the above given equation is explained here. First when A wants to communicates with node D at this time according to table 8 the shortest depth is selected so A will be placed at $Xi, S, t \times Ptx$ so the energy consumption at node A will be considered. After that at this time node A will select the shortest depth node which is node F then it will become the receiving node and the energy consumption for request to send signal and clear to send signal should also be considered. The energy consumption calculation is presented by $Prts + Pcts$ at every sender and receiver node involved in the communication between node A and node F, so the node B which is also neighbour of the node A so it will become the idle listening node, hence the total number of nodes involved in the receiving and idle listening are two, node F and node B respectively. Here node F's energy consumption is presented by $Xi, R, t \times Prcv$ and the energy consumption at node B is presented by the term $Xi, L, t \times Ptx$. There is no need to calculate the energy consumption at node B because according to propose model that node B is scheduled to be sent in to sleep mode for a specific time threshold if rts and cts signal is received successfully at node A from node F. Thus the idle listening mode is merged with the sleeping mode, so only node A who is transmitter and node F who is receiver only two node's energy consumptions are needed to be considered. In the next step when node F receives the data successfully and send acknowledgement to node A and receives the confirmation from node A then node F will be placed in to transmission mode presented by $Xi, S, t \times Ptx$ and according to equation the node E will be coming in receiving state mode presented by $Xi, R, t \times Prcv$. At this time the power consumption between node F and node E for sending rts and cts signal must also be considered and it is presented by $Prts + Pcts$. Now the value of k will be one in this case because there is no idle listening mode because node E has only two neighbors one is node F and another is node D as shown in figure 1.Thus there are only two nodes one is sender and another is receiver will be present in the consumption of energy only when any node in WSN has only two neighbors, and if any node has more than two neighbors in that case the consumption will follow the proposed equation.

The time threshold must be set so that till the finish of the transmission every nodes will be awaken up in considerable negligible time. The power consumption during various stages is explained below in a brief manner.

Ptx = total power consumption at nodes involved in communication in static network where no node movement is done.

$$Ptx = k \times (Prts + Pcts + Ptrans + Pack + Packconf)$$

Where $k$ = number of nodes involved during the communication from sender node to receiver node, $(k < j)$ where j is total number of nodes existing inside the clusters. Prts = power consumption in sending request to send signal, Pcts = power consumption in receiving clear to send signal, Ptrans = power consumption in transmitting the packets, Pack = power consumption in getting acknowledgment signal, Packconf = power consumption in sending the confirmation of acknowledgment signal on reception.

In above equation Prts + Pcts term is the power consumption for making sure that there is no link failure in the network. Ptrans term is used for data packet transmission and Pack + Packconf term is used for 3 way handshaking mechanism.

Ptx = total power consumption at nodes involved in communication in dynamic sensor network when node movement is done.

Case (a) When node wants to leave the cluster.

$$Ptx = [\{k \times (Prts + Pcts + Ptrans + Pack + Packconf)\} + (v \times Pnode) + PtransCH + PrcvCH + (n \times Prcv)]$$

Where $v$ = number of nodes that want to move outside the cluster, Pnode = power consumption at node during the transmission of bit 0 message to the cluster head, PtransCH = power consumption at cluster head during broadcasting of the updated information to the cluster nodes, PrcvCH = power consumption at cluster head in receiving bit 0 message. $n$ = number of nodes receiving the broadcasted message from the cluster head, Prcv = power consumption at nodes during reception of the broadcasted message for the updation of their table.

Case (b) When node wants to add in to the cluster

$$Ptx = [\{k \times (Prts + Pcts + Ptrans + Pack + Packconf)\} + (h \times Ptr) + PtransCH + PrcvCH + (b \times Pcomm)]$$

Where $h$ = number of nodes which transmit the newly added node's information to the cluster head, Ptr = power consumption during the transmission of information about newly added node to the cluster head by the nodes, $b$ = number of nodes whose range overlaps with the newly added node inside the cluster, Pcomm = power consumption during the communication among the previously existing cluster nodes and newly added cluster node.

## 4. ADVANTAGES

In the proposed model the link failure is getting notified to the sender node by not receiving the cts (clear to send) signal from the receiver node, so the sender node will be selecting the other possible best route for transmitting the data frames even if the possible route is having the longest depth. With the help of this if there is a communication or a link failure between nodes as well as in channel then also alternate route will be helping to send the data frames very smoothly. If other possible paths are also showing link failures then definitely there would be no chance to send the frames because for sending the frames there must be at least one healthy path available without failure. The main sender node and final destination node may or may not be synchronized with each other. In the proposed case the main sender node is A who wants to send data frames to node D but when node D receives the data frames at that time node A will be sleeping probably because after transmitting data frames to node F by node A, it is the headache of node F to pass the data frames further in the network till node D. Suppose, node A wants to send data frames to node F in that case the node A is main sender node and node F is final destination node. In this scenario it is obvious that both nodes must be synchronized with each other.If there is no link failure and every nodes are doing communication process smoothly then there will be guarantee for any node that the message sent by it will definitely reach to the intended node successfully even if that node sleeps.

The proposed model ignores the energy consumption of nodes that are in idle listening state by putting them in to sleep mode, so every time there are only two nodes' energy consumption need to be considered at each and every stage.

The proposed model doesn't allow interference of one node as main sender for sending rts signal to any other existing node in the network because there will be no new communication can possible till previous communication ends with in specific time because of allocated specific time threshold to every node priority wise. The energy consumption presented by $Prts + Pcts$ avoids any interference by other node while the previous communication is being done.

Dynamic priority changing mechanism provides chance to every node for becoming highest priority node once so they can be main sender of the network and can communicate with any node by sending rts signal to the desired node for further complete destination node till its message get conveyed to the destination node.

## 5. CONCLUSION

This model concludes that the energy wasted at the nodes while idle listening is saved by putting them in to sleep mode. Every time there would be only two nodes whose energy consumption needs to be considered. This model helps for efficient energy scheduling when there are a lot of clusters available inside wireless sensor network and inside these clusters there are a lot of nodes in the different clusters. Adjacency matrix formed at controller node provides a fruitful way in forming the complete information about the different nodes for a specific cluster and passes that information to the different cluster according to the number of the clusters for the static network when there is no movement of a node is allowed. For a static network the power consumption is defined by the equation $Ptx = k \times (Prts + Pcts + Ptrans + Pack + Packconf)$ in which the value of k is always less than the total number of existing nodes inside the cluster. In the case of the dynamic sensor network where the node movement is allowed there are two cases which must be included during the calculation of the power consumption. In the case of the dynamic sensor network where the movement of the node is allowed the power consumption will be considered in the case of when the node is leaving the cluster will be $Ptx = [\{k \times (Prts + Pcts + Ptrans + Pack + Packconf)\} + (v \times Pnode) + PtransCH + PrcvCH + (n \times Prcv)]$ . In the case of when the node is being added into the cluster the energy consumption will be $Ptx = [\{k \times (Prts + Pcts + Ptrans + Pack + Packconf)\} + (h \times Ptr) + PtransCH + PrcvCH + (b \times Pcomm)]$ . In the given model there is provision for both the cases first is multiple clusters can be connected to the single router and individual cluster can be connected to their individual router to communicate with the controller node. This helps clusters to communicate very smoothly in a heavy load condition. When every router in side the network becomes ready to send the information to the controller node, and if there will be one controller node existing in the network, it may possible to occur a bottleneck situation for the controller node. To handle the load condition occur at the controller node there should be provision for more than one controller nodes inside the wireless sensor network those can manage the different clusters in heavy traffic condition, form the adjacency matrix and pass the information to the routers cluster wise as a future scope. However the model proposed in this paper ensures the data delivery with energy efficient scheduling of nodes.

## Authors

Shweta Singh received her B.TECH. degree in Computer Science Engineering from Subharti Institute of Technology and Engineering in 2010.Currently she is pursuing M.TECH degree in Computer Science and Engineering at Jaypee University of Information Technology, Waknaghat, Distt. Solan- 173215. Her areas of interest are Theory of computation, Advanced Algorithms, Software Engineering, Computer Networks**.** Currently she is working her thesis work on Energy Efficient Scheduling Maximizing energy saving in Wireless Sensor Networks.

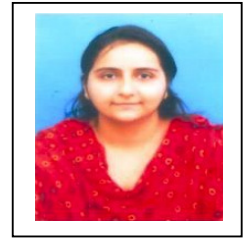

Ravindara Bhatt is currently working in Computer Science and Engineering Dept as a faculty  at  Jaypee University of Information Technology, Waknaghat, Distt. Solan- 173215 and he is currently pursuing Ph.D from IIT Kharagpur. He received his M.Tech degree in Information Technology from Guru Gobind Singh Indraprastha University, Delhi in 2005. He got his Diploma in Advanced Computing from C-DAC in 2001. He received his B.E degree in Electronics & Communication from H. N. B. Garhwal University, Uttaranchal in 1996. His areas of Interest are Cryptography & Network Security, Image Processing, Object oriented programming.

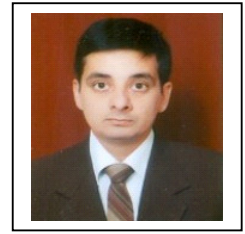